\input{aipcheck}

\documentclass[
    ,final            
  ,numberedheadings 
  ]
  {aipproc}

\layoutstyle{6x9}

\begin{document}

\title{TeV Scale Spontaneous R-Parity Violation
\footnote{Based on a plenary talk given by P.F.P. at PPC09, OU, May 2009 and talk given by S.S. 
at SUSY09, Northeastern Univ., Boston, June 2009.}}


\author{Pavel Fileviez P{\'e}rez}{
  address={Department of Physics, University of Wisconsin, Madison, WI 53706, USA}
}

\author{Sogee Spinner}{
  address={Department of Physics, University of Wisconsin, Madison, WI 53706, USA}
}

\begin{abstract}
Understanding the origin or absence of the $R$-parity violating interactions in the 
minimal supersymmetric Standard Model is a vital and open issue.
Here we show that in the minimal $B-L$ models, $R$-parity and $B-L$
are spontaneously broken at the TeV scale.  We also briefly discuss the
phenomenological and cosmological aspects of these scenarios.
\end{abstract}

\maketitle


\section{Introduction}
Supersymmetry (SUSY) has captured the imagination of many of the practitioners 
of our field due to the elegant solution to the hierarchy problem and rich phenomenology.  
An open issue in SUSY is the origin of
lepton (L) and baryon (B) number violating interactions:
$$\hat L \ \hat H_u,
\quad \hat L \ \hat L \ \hat E^C,
\quad \hat Q \ \hat L \ \hat D^C,
\quad \hat U^C \ \hat D^C \ \hat D^C.$$
%
%
Since the phenomenological and cosmological aspects of the MSSM hinges on the
presence or absence of these interactions, it is crucial to understand the status 
of the so-called $R$-parity symmetry. $R$-parity is defined as $R = (-1)^{3\left(B-L\right) + 2S} = (-1)^{2S} M$,  where $M$ is called matter parity and $B$, $L$ and $S$ stand for baryon number,
lepton number and spin, respectively.  Conservation of $R$-parity forbids the dangerous
processes due to the operators listed above, \textit{e.g.} the dimension four contributions for proton decay, and guarantees the
stability of the lightest supersymmetric particle (LSP), a good candidate for the cold
dark matter of the Universe.  Although even if $R$-Parity is broken, 
the gravitino still makes a good dark matter candidate.  Therefore, understanding
the origin or absence of $R$-parity conservation
is key to appreciating the phenomenological and cosmological implications of the SUSY program.
This is the main focus of our letter which
discusses our findings published in Refs.~\cite{S1},~\cite{S2},~\cite{S3} and~\cite{S4}.

Examining the definition of $M$-parity leads one to imagine several scenarios related to $B-L$: 
i) if $B-L$ is conserved, $M$ is always conserved. ii) If $B-L$ is broken by an even number, 
$M$ is conserved, and iii) if $B-L$ is broken by an odd number, $M$-Parity is violated. These
scenarios can be realized in either global or local $B-L$ symmetries.  Global $B-L$ has been
studied in the past~\cite{global-B-L} and  necessitates facing the Majoron problem~\cite{Majoron}, 
while a local, $R$-parity conserving approach, was investigated in a systematic way in Ref.~\cite{Martin}. 
Here we focus our attention on the latter case of local $B-L$ symmetries and show that \textit{the minimal $B-L$ model always requires the spontaneous breaking of $R$-Parity}.

We have investigated this idea for the origin of the $R$-parity violating interactions in four different scenarios~\cite{S1,S2,S3,S4} where the electroweak gauge group at the TeV scale is
\begin{itemize}
\item a) \ $SU(2)_L \bigotimes U(1)_Y \bigotimes U(1)_{B-L}$, 
\  \  \  \  b)  \ $SU(2)_L \bigotimes U(1)_Y \bigotimes U(1)_{\left( a \ Y \ + \ b \ (B-L) \right)}$,
\item c) \ $SU(2)_R \bigotimes SU(2)_L \bigotimes U(1)_{B-L}$, \  \  \  \  d) \ $SU(2)_L \bigotimes U(1)_{I_3^R} \bigotimes U(1)_{B-L}$.
\end{itemize}
In all these theories $B-L$ is part of the local symmetry and the minimal matter content
includes only the MSSM states plus three generations of right-handed neutrinos (required
by anomaly cancellation).  If no new states are added in by hand, the only way to break the gauge symmetry down to the SM gauge symmetry must also break $R$-parity spontaneously. 
This is achieved through the vacuum expectation value of the right-handed sneutrino.  The resulting
theory is a simple TeV scale extension of the MSSM, with the following attractive properties:

\begin{enumerate}

\item A mechanism for spontaneous $R$-parity violation (SRpV), which generates only bilinear terms.

\item The $R$-parity and $B-L$ violating scales are defined by the SUSY mass scale (TeV scale).

\item Rapid proton decay is avoided since Baryon number is preserved at the renormalizable level.

\item Neutrino masses are generated at tree level through $R$-parity violation and the type I seesaw mechanism.

\item Testability at the LHC through the $Z'$ properties and $R$-parity violating decays.

\item The gravitino as a viable dark matter candidate.

\end{enumerate}
In the next section we discuss in detail the mechanism for spontaneous $R$-parity violation in these scenarios.
\section{Local $B-L$ Symmetry and SRpV}
\label{SRPV}
In the minimal $B-L$ extension of the MSSM the gauge group of the electroweak sector 
is based on $G_{B-L}=SU(2)_L \bigotimes U(1)_Y \bigotimes U(1)_{B-L}$. The 
particle content and its $G_{B-L}$ charges are given by
\begin{eqnarray}
\hat{Q}^T &=& \left(\hat{U}, \hat{D}\right) \sim (2,1/3, 1/3), 
\;\;\;\; \hat{U}^C \sim (1, -4/3, -1/3),\nonumber \\
\hat{D}^C& \sim & (1,2/3,-1/3),
 \qquad \hat{L}^T = \left( \hat{N}, \hat{E} \right) \sim (2,-1,-1),\nonumber \\
 \hat{E}^C &\sim & (1,2,1), \qquad \qquad \hat{N}^C \sim (1,0,1),
 \label{matt}
 \end{eqnarray}
and as we know the MSSM Higgses have no $B-L$ charge. 
In this case the most general superpotential is
\begin{eqnarray}
 	\mathcal{W} & = & \mathcal{W}_\mathrm{MSSM} \ + \ Y_\nu \ \hat L \ \hat H_u \ \hat N^C,
\end{eqnarray}
with
\begin{eqnarray}
	\mathcal{W}_\mathrm{MSSM} & = & 
	Y_u \ \hat Q \ \hat H_u \ \hat U^C
\ + \ Y_d \ \hat Q \  \hat H_d \  \hat D^C
\ + \ Y_e \  \hat L \  \hat H_d \  \hat E^C
\ + \  \mu \ \hat H_u  \ \hat H_d.
 \end{eqnarray}
The soft SUSY breaking potential contains the terms
\begin{eqnarray}
	\nonumber
	V_{soft} & \supset & M_{\tilde N^C}^2 \ \vert \tilde{N}^C\vert^2
		\ + \ M_{\tilde L}^2 \  \vert \tilde L\vert^2 
		\ + \ M_{\tilde E^C}^2 \  \vert \tilde E^C \vert^2 
		\ + \ m_{H_u}^2 \  \vert H_U \vert^2
		\ + \ m_{H_d}^2 \  \vert H_D \vert^2 
	\nonumber
	\\ &+& 
	\left(
		A_\nu \ \tilde{L} \ H_u \ \tilde{N}^C
		 \ + \ \mathrm{h.c.}
	\right) .
\end{eqnarray}
At this point, it may seem like an extra field is needed in order to break the gauge symmetry to the SM.
However, the minimal model already contains the necessary ingredient, the right-handed neutrino.  Therefore \textit{in the minimal $B-L$ model $R$-parity is spontaneously broken}. 
We proceed by investigating the scalar potential.  Once one generation of sneutrinos, $\tilde \nu$ and $\tilde \nu^C$ and the Higgses acquire vacuum expectation values 
(VEVs) $v_L, v_R$ and $v_{u,d}$ (with $v_u^2 + v_d^2 = v^2$) respectively, the components of the scalar potential reads
as
\begin{eqnarray}
	\left<V_F\right> & = &
		\frac{1}{4} Y^2_\nu
		\left(
			v_R^2 \ v_u^2
			+ v_R^2 \ v_L^2
			+ v_L^2 \ v_u^2
		\right)
		+ \frac{1}{2} \mu^2 v^2
		- \frac{1}{2} Y_\nu \ \mu \ v_d \ v_L \ v_R,
	\\
	\left<V_\mathrm{soft}\right> & = &
		\frac{1}{2} \left(
			M_{\tilde L}^2 \ v_L^2
			\ + \ M_{\tilde N^C}^2 \ v_R^2
		\right)
		+ \frac{1}{\sqrt 2} A_\nu \ v_u \ v_L \ v_R,
	\\
	\left<V_D\right> & = &
		 \frac{1}{32} g_{BL}^2 \left(v_R^2 - v_L^2 \right)^2.
\end{eqnarray}
where we have ignored the pure MSSM Higgs contributions.

Minimizing in the limit $v_R \gg v_u, v_d \gg v_L$
\begin{eqnarray}
	\label{vR}
	v_R & = & \sqrt{\frac{-8 M^2_{\tilde N^C}}{g_{BL}^2}},
	\  \  \  \ 
	v_L  = \frac{v_R \ B_\nu}{M^2_{\tilde L} - \frac{1}{8} g^2_{BL} v_R^2},
\end{eqnarray}
and $B_\nu = \frac{1}{\sqrt 2} \left(Y_\nu \mu v_d - A_\nu v_u\right)$.  Notice that the 
expression for the right-handed sneutrino VEV is extremely simple in this case and resembles
that of the SM.  It also indicates that the right-handed sneutrino needs a negative soft mass squared
to achieve symmetry breaking.  While this is possible through some modifications of the variety of popular
SUSY breaking mechanisms available, it is important to note that it is also possible through the inclusion of
Fayet-Iliopoulos terms for $G_{B-L}$.  Regardless, it is crucial to mention that the $R$-parity breaking scale 
is defined by the SUSY breaking scale.

\textit{$R$-parity Violation}: After symmetry breaking, the effective MSSM-like theory will contain $R$-parity violating bilinear 
terms, all of which will be defined by two VEVs: $v_L$ and $v_R$, where $v_R \ \gg \ v_L$. For example, the $Y_\nu \ \hat{L}^T \ i \sigma_2 \ \hat{H}_u \ \hat{N}^C$
term in the superpotential, leads to $Y_\nu \ l \ \tilde{H}_u \ v_R / \sqrt{2}$ 
and $Y_\nu  \ \tilde{H}_u^0  \ \nu^C \ v_L/ \sqrt{2}$.  Other generated bilinears, such as mixing between
the neutrinos and neutral gauginos, will be proportional to $v_L$ and will therefore
be quite small, while the mixing between the $B-L$ gaugino and right-handed neutrino
will be proportional to $v_R$, but will not lead to any dangerous low energy consequences.
Notice that only bi-linear $R$-parity violating terms which violate lepton number are generated. Therefore, there are no contributions to proton decay at the renormalizable level. See~\cite{S3,review} for the discussion of the relevant proton decay  contributions coming from higher-dimensional operators.

Schematically, the neutrino mass in these types of models have two contributions given by
\begin{eqnarray}
M_\nu & = & \frac{1}{2} \ Y_\nu \ M^{-1}_{\nu^C} \ Y_\nu \ v_u^2 \ + \ m^T \ M_{\tilde \chi^0}^{-1} \ m,
\end{eqnarray}
where the first is a type I seesaw contribution and the second comes from the $R$-parity induced
mixings of the neutrinos and neutralinos.  Since in both contributions the "seesaw scale" is the TeV, 
the corresponding Yukawa couplings, $Y_\nu$ must be quite small on the order of $10^{-6}$ or lower.  
See Refs.~\cite{S1},~\cite{S2},~\cite{S3} and~\cite{S4} for details.

As was the case for the neutrinos, which mixed with the neutralinos, the charged leptons and 
sleptons will mix with the charginos and Higgses respectively.  However, these mixings are
usually proportional to the neutrino mass parameters and are therefore suppressed and do not
significantly affect the spectrum.  This means that the  only deviations from the MSSM spectrum
will be the 
CP-even right-handed sneutrino, corresponding to the $B-L$ breaking VEV (degenerate 
with the $Z'$ gauge boson) and $D$-term contributions to the sfermion masses.  Any important
deviations from this general discussion will be pointed out below where specific
larger gauge groups (containing $B-L$) will be considered.
\\
\textit{\underline{Beyond The Minimal $B-L$ Extension of the MSSM}}:
\begin{itemize}

\item $G_X=SU(2)_L \bigotimes U(1)_Y \bigotimes U(1)_{X}$: 
The mechanism discussed in the previous section can be applied in a different scenario where 
$B-L$ is just part of the gauge symmetry. This is the case of a $U(1)_X$ extension of the MSSM 
where X is a linear combination of weak hypercharge and $B-L$, $X = a Y + b (B-L)$. 
The results of our investigation have been published in Ref.~\cite{S3}. These scenarios are 
better motivated if one takes into account seriously the idea of Grand Unification 
because the case $a=1, \ b = -5/4$ corresponds to the  embedding in $SO(10)$.

This scenario differs from the previous one discussed before due to the non-zero $X$
charge of the MSSM Higgses, and of course the minimization conditions are different.  
Doing the same exercise as before we find that 
\begin{equation}
		v_R = \sqrt{\frac{-8 M^2_{\tilde N^C} \ - \ a \ b \ g_X^2 \  \left(v_u^2-v_d^2\right)}{g_{X}^2 \ b^2}},
\end{equation}
where the presence of the second term in the numerator could aid $U(1)_X$ symmetry breaking,
even in the case of a non-negative $M^2_{\tilde N^C}$ although this parameter would need to be very
small in that case. In this theory the resulting spectrum is quite different. For example, 
sfermion masses will have the following contributions from $X$ $D$-terms:
\begin{equation}
	\delta m_\phi^2 = \frac{1}{8} \  X\left(\phi\right) \left[b v_R^2 + a \left(v_u^2 - v_d^2\right)\right],
\end{equation}
where we have ignored contributions from $v_L$. The full spectrum of these models 
have been investigated in Ref.~\cite{S3} and we invite the reader to study this paper.
\\

\item $G_{LR}=SU(2)_L \bigotimes SU(2)_R \bigotimes U(1)_{B-L}$: Even more interesting 
than the Abelian case is the utilization of this mechanism in the context of left-right 
symmetric models~\cite{S1}, in which the right-handed isospin group 
and $B-L$ break into the SM hypercharge group. 
For $R$-parity conservation in this context, an 
involved Higgs sector is required~\cite{LR-RpC}.  Once again though, symmetry breaking and
SRpV can be achieved with just the minimal content.  Unfortunately, $G_{LR}$ models need two Higgs bidoublets in order to have a consistent relation between quark 
masses.  This leads to severe constraints coming from flavour violation. \\

\item $G_{I^R_3}=SU(2)_L \bigotimes U(1)_{I^R_3} \bigotimes U(1)_{B-L}$: An interesting 
alternative to the previous scenario arises when $G_{LR}$ breaks to $G_{I^R_3}$ at some high-scale 
leaving the TeV scale gauge symmetry defined by $G_{I^R_3}$. 
In this case it is the combination of the two Abelian groups which breaks into the SM hypercharge.
Here, the VEV for the right-handed sneutrinos read as~\cite{S4}
\begin{equation}
	v_R  = \sqrt
		{
			\frac{-8 M_{\tilde{N}^C}^2 + g_R^2 \left(v_u^2 - v_d^2\right)}
			{g_R^2 + g_{BL}^2}
		}
\end{equation}
where $g_R$ is the $U(1)_{I^R_3}$ gauge coupling. 
Here again the $D$-term contributions to the soft 
masses are different since they do not include the hypercharge contributions 
that exist in the MSSM.  Therefore, in addition to the typical $SU(2)_L$ 
contributions, one also has
\begin{equation}
	\delta m^2_\phi = \frac{1}{8} BL\left(\phi \right) v_R^2
		- \frac{1}{8} I^R_3\left(\phi\right) \left(v_R^2 + v_d^2 - v_u^2\right),
\end{equation}
where $BL\left(\phi \right)$ is the $B-L$ charge of $\phi$.

\end{itemize}

\textit{\underline{Collider Signals}}:
As a consequence of $R$-parity violation, the lightest neutralino will be
unstable and will decay via lepton number violating interactions. These 
interactions will also exist for the charginos and the new gauge boson and can contribute 
to an interesting scenario which would arise when the sneutrino is the NLSP 
and the gravitino the LSP. In this case the $Z'$ allows for a new production 
mechanism for sleptons at the LHC:
$$ pp \ \to \ Z' \ \to \  \tilde{\nu} \tilde{\nu}^* \ \to \  e^+_i \ e^-_j \ e^+_k \ e^-_l.$$
Since the sneutrino is the NLSP, it must decay through $R$-parity violating interactions with the
following possibilities: $\tilde{\nu} \ \to \nu \nu, e^+_i e^-_j$.  This will lead to channels with four
leptons in the final state such as : $eeee$, $e \mu \mu \mu$, $e e \mu \mu$, $e e e \mu$, $\mu \mu \mu
\mu$ and also with several tau's. With such spectacular signals, one could 
test the existence of $R$-parity violation and lepton number violation. 
Distinguishing between the different scenarios discussed above 
can be accomplished by studying the properties of the $Z'$ and the different $R$-parity violating decays.
\section{SUMMARY}
The possibility of understanding the origin or absence of the $R$-parity violating 
interactions in low energy SUSY models has been discussed. 
Specifically, we have pointed out that in minimal models where $B-L$ is part 
of the local symmetry $R$-parity should be spontaneously broken, along with $B-L$, 
at the TeV scale.  We have 
discussed this mechanism for spontaneous $R$-parity violation in different 
extensions of the MSSM, pointing out ways of distinguishing between these at colliders.
In these scenarios the gravitino can be the cold dark matter candidate 
and channels with multi-leptons are crucial for testing this idea at the LHC.

\begin{theacknowledgments}
It is pleasure to thank our collaborators V. Barger and L. Everett 
for their contributions to the work presented here. The 
work of P.F.P. was supported in part by the U.S. Department 
of Energy contract No. DE-FG02-08ER41531 and in part
by the Wisconsin Alumni Research Foundation. 
S.S. is supported in part by the U.S. Department 
of Energy under grant No. DE-FG02-95ER40896,
and by the Wisconsin Alumni Research Foundation.
Last, but not least, we would like to thank Espresso 
Royale for hospitality.
\end{theacknowledgments}





\begin{thebibliography}{9}

\bibitem{S1}
  P.~Fileviez P\'erez and S.~Spinner,
  Phys.\ Lett.\  B {\bf 673}, 251 (2009).

\bibitem{S2}
  V.~Barger, P.~Fileviez P\'erez and S.~Spinner,
  Phys.\ Rev.\ Lett.\  {\bf 102}, 181802 (2009).

\bibitem{S3}
  P.~Fileviez P\'erez and S.~Spinner,
  Phys.\ Rev.\  D {\bf 80}, 015004 (2009).

\bibitem{S4}
  L.~L.~Everett, P.~Fileviez~P\'erez and S.~Spinner,
  Phys.\ Rev.\  D {\bf 80} (2009) 055007.

\bibitem{global-B-L}
  A.~Masiero and J.~W.~F.~Valle,
  Phys.\ Lett.\  B {\bf 251} (1990) 273;
  J.~C.~Romao, C.~A.~Santos and J.~W.~F.~Valle,
  Phys.\ Lett.\  B {\bf 288} (1992) 311;
  M.~Shiraishi, I.~Umemura and K.~Yamamoto,
  Phys.\ Lett.\  B {\bf 313} (1993) 89;
  G.~F.~Giudice \textit{et al},
  Nucl.\ Phys.\  B {\bf 396}, 243 (1993).

\bibitem{Majoron}
  G.~B.~Gelmini and M.~Roncadelli,
  Phys.\ Lett.\  B {\bf 99} (1981) 411;
  C.~S.~Aulakh and R.~N.~Mohapatra,
  Phys.\ Lett.\  B {\bf 119} (1982) 136;
  Y.~Chikashige, R.~N.~Mohapatra and R.~D.~Peccei,
  Phys.\ Lett.\  B {\bf 98} (1981) 265.

\bibitem{Martin}
See for example: 
R.~N.~Mohapatra,
 Phys.\ Rev.\  D {\bf 34}, 3457 (1986);
S.~P.~Martin,
  Phys.\ Rev.\  D {\bf 46} (1992) 2769;
  Phys.\ Rev.\  D {\bf 54} (1996) 2340.

\bibitem{review}
  P.~Nath and P.~Fileviez P\'erez,
  Phys.\ Rept.\  {\bf 441} (2007) 191;
  B.~Bajc, P.~Fileviez P\'erez and G.~Senjanovic,
  Phys.\ Rev.\  D {\bf 66} (2002) 075005;
  arXiv:hep-ph/0210374.

\bibitem{LR-RpC}
  C.~S.~Aulakh, A.~Melfo and G.~Senjanovic,
  Phys.\ Rev.\  D {\bf 57} (1998) 4174;
  R.~Kuchimanchi and R.~N.~Mohapatra,
  Phys.\ Rev.\  D {\bf 48} (1993) 4352;
  K.~S.~Babu and R.~N.~Mohapatra,
  Phys.\ Lett.\  B {\bf 668} (2008) 404
\end{thebibliography}


\end{document}